# Department Management System for Departments of Sri Lankan Universities

Thuseethan, S.

**Abstract:** This paper proposes a new method of data handling as well as implementation of management system for an academic department. Management system is a proven framework for managing and continually improving the organization's policies, procedures and processes. Department of Sri Lankan Universities is a division within a faculty comprising one subject area or a number of related subject areas. The time consumption and error rate of producing information is extremely high. This paper describes how to do the data handling in a department of Sri Lankan Universities using the management system. The efficiency of the management system measured with the amount of data handled by the department. Experiment results shows that this method of data handling increase the efficiency of the department in terms of processing information and the reduction of risks involved with the department activities.

----------------------- ◆ -----------------------

## 1. INTRODUCTION

We are in the field of Information Technology, with an importance beyond that of providing basis for the use of computer in day-to-day work. Computer is a machine which made by man and now he is trying to give artificial brain to go hand in hand with the human mind. Presently there is a big battle between the computer and the man mind. As far as my concerned that should not be happen. Both the human brain and the computer should go in parallel. It would be beneficial if the technology can mingle with the management of department.

This research helps to organize and administrate departmental works and also to improve the quality of a routine works, especially under the situation of no such system is been used in this field in Sri Lanka. Nowadays the world grows we need to be fast, fluid and flexible (Achchuthan and Nimalathasan, 2012). We need to develop these capabilities to survive. In this manner educational organizations are also wish to follow this kind of environmental need (Sivarajah and Achchuthan, 2013). Universities are maintaining University Management System (UMS) to ensure their capabilities.

University Management System (UMS) automates all the university transactions, internal work flow procedures of a university as well as the interaction with the students and the instructors. University Management System (UMS) combines a suite of applications specifically designed to bring together People, Process and Technology towards improving Relevance and Quality of higher education.

Department Manager Software for Universities is a powerful financial and personnel management software tool designed for administrators of large departments, research institutes and colleges. These types of software are effective, timesaving solution for managing department and they are help to manage special commitments and information unique to the organization and bring flexibility to the data management tasks.

The universities under the University Grant Commission (UGC) of Sri Lanka are maintaining several University Management Systems (UMS). But these systems are not a generalized, in order to computerize the campus, as the first step I developed Department management system for the Department of Physical Science, Vavuniya Campus, Sri Lanka which provides the up-to-date information of the entire department.

## 2. PROBLEM DOMAIN AND AIM

The Management Information System (MIS) is a concept of the last decade or two, involve a broad and complex topic. It has been understood and described in a number ways. It is also known as the Information System, the Information and Decision System, the Computer- based information System. Management Information System (MIS) provides information for the managerial activities in an organization (G.Satyanarayana reddy,Rallabandi Srinivasu,Srikanth Reddy Rikkula, Vuda Sreenivasa Rao, 2009).

Although there are many education based departments, there are less computerized management systems to manage such departments in Sri Lanka especially in northern part. Considering the fact that there are around fifty departments in Sri Lankan universities and University Grant Commission is the main organization which has the right to co-ordinate the functions in these departments.

At present, most of the departments organizing and managing activities have been done manually. According to the situation of current analysis departments faces difficulties when they managing the routine works that has been done by them. Some of them are;
- Data security and reliability
- Time consumption during their process.
- High cost activities.
- Data Validations
- Difficult to produce analytical reports
- Hard to fill data entry forms.
- No user friendliness.
- It takes too much time to get the details of any of the student and employee, book etc.
- Need a lot of paper work (registers).
- Entry mistake are another problem for the system and have drawback of accuracy of result obtained.

A computer based information system is usually needed for the following purposes.
- Greater Processing Speed: Using computers inherent ability to calculate, sort, retrieve data with greater speed than that of the human doing we can get results in less time. Dot Net guarantees from

Author: Thuseethan, S., Lecturer in Computer Science, Department of Computing and Information Systems, Sabaragamuwa University of Sri Lanka, Sri Lanka.
Email: thuseethan@gmail.com

the faster query processing thus we are satisfied with Dot Net itself supporting in the direction.
- Better Accuracy and Improvement Consistency: The computer carries out computing steps including arithmetic accurately and consistently from which really human is escaped which yields more fatigue and boredom.
- Cost Reduction: Using computerization we can do the required operations with lower cost than any other methods. Hence by computerization we can reduce the cost drastically.

Therefore through this research I endeavor to improve and maintain the standards, quality and efficiently of the department. The system also effective in cost wise and human resource wise. In addition all the department heads, assistants, staffs, students and University Grant Commission of Sri Lanka will benefit immensely from this system.

## 3. METHODOLOGY

Very comprehensive requirement analysis was done, in order to get a very good understanding about the problem domain. During the design phase not only the theories but also the functional and nonfunctional requirements were kept in mind. Due to that some design concept were slightly violated in order to achieve better outcome with respect to the functionality. Implementation was done according to the design and it was done according to the object oriented techniques and followed three-tier architecture. Although it was put full effort to implement in proper manner, at the end it was realized that further optimization and fine tuning work can be carried out. User interfaces were designed after considering lot of factors in such a way that, end users will get motivated to use this system. Also it was a big challenge to find out better data set for testing purposes.

Functionalities like some tie-breaking systems were not able to test properly using the data which had with department. So some planned data was used in order to test some rarely occurring situations. As a whole the implemented system was successful, since it achieved all the planned functional and non-functional requirements as well as some additional features. So, as an improvement to the developed system, facilities can be provided to manage the Department of Physical Science activities. This will added more value to the developed system. Further it covers almost all the areas of Department of Physical Science administration in DPS. In addition, developed web application was provided facilities like; retrieving relevant information.

## 4. TOOLS USED

A set of software was selected to implement the system, by considering factors like support for object oriented implementation, facilitate for high performance and user friendly system, compatibility with each other and the functional and non-functional requirements of the intended system. Microsoft Visual Studio .NET Enterprise Edition 2008 was used to implement the system in Microsoft Visual C# .NET language the Windows platform. My choice is based on the reliable programming environment offered by C# and the highly cleared syntax of this programming language as well. In order to implement data layer Microsoft SQL Server 2005 get involved. Transact-SQL (TSQL) was used as a development language in the database layer and SQL Query Analyzer was employed as a development tool. Further Crystal Reports 9.0 and the tool HTML Help Workshop were utilized to implement reports and help facility, respectively. In addition Macromedia Dreamweaver MX was used to design the web pages and also deal with HTML in web development.

Finally, developed software used as proof of concept to analyze the speed up difference in academic department.

## 5. RESULT AND DISCUSSION

The results indicate that a huge amount of improvement in the departments of Sri Lankan universities with the use of department management system. So far I have developed application software called "DPS Management Information System" to maintain the Student details, Staff details and inventory control of the Department of Physical Science. The following graph depicted the speedup with respect to amount of data.

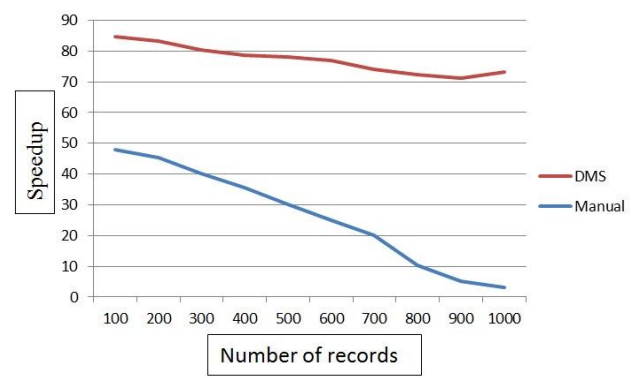

The overall speedup is more than 70% with the use of management system. The maximum speedup achieved by manual work and department management system is 47.8% and 84.3% respectively. The minimum speedup achieved by both manual work and department management system are 3.4% and 73.1%. This maximum speedup achieved while handling small amount record and minimum is achieved while handling huge amount of records. Data retrieval is also easy and fast. This also restricts the users to enter invalid data and reduces the burden on the management.

The error rate is 0.01% in doing the routine works with the help of department management systems while error percentage is more than 12% in manual works.

## 6. CONCLUSION

The overall speedup is considerably higher than the manual work. The management system enhances the functionalities of the routine works of the department in a number of ways. The computerization helps the users a lot to minimize the working time with ease. The department staffs get information in desired manner. Data retrieval is also easy and fast. This also restricts the users to enter invalid data and reduces the burden on the department. The data maintaining has been made quite simple such as searching of records and records maintenance. The error rate reduced dramatically with the use of department management system.

## 7. REFERENCES

[1] Treloar, A. (2005). Developing an Information Management Strategy for Monash University'. *Proceedings of Educause AustralAsia 2005*.

[2] Achchuthan, S., & Nimalathasan, B. (2012). Level of entrepreneurial intention of the management undergraduates in the university of jaffna, Sri Lanka: Scholars and undergraduates perspective. *ACADEMICIA: An International Multidisciplinary Research Journal, 2(10), 24-42.*

[3] Haag, S., Cummings, M., & Dawkins, J. (1998). Management information systems. *Multimedia systems, 279, 280-297.*

[4] Lucey, T. (2005). Management information systems. Cengage Learning EMEA.

[5] Laudon, K. C., & Laudon, J. P. (2011). Essentials of management information systems. Upper Saddle River: Pearson.

[6] McLeod Jr, R., & Schell, G. (2001). Management Information Systems 8/E.Chapter-17" Marketing Information System" published in.

[7] Reddy, G. S., Srinivasu, R., Rikkula, S. R., & Rao, V. S. (2009). Management information system to help managers for providing decision making in an organization. *International Journal of Reviews in Computing, 1-6.*

[8] Sivarajah, K., & Achchuthan, S. (2013). Entrepreneurial Intention among Undergraduates: Review of Literature. *European Journal of Business and Management*, *5*(5), 172-186.